\documentclass[aps,prl,twocolumn,showpacs,superscriptaddress,groupedaddress]{revtex4}  
\usepackage{graphicx}  
\usepackage{bm}        
\usepackage{amssymb}   
\usepackage{verbatim}   
\usepackage{subfigure}
\begin{document}
\title{Electron rescattering in strong-field photodetachment of F$^-$}

\author{O. Hassouneh} 
\author{S. Law}
\author{S. F. C. Shearer}
\author{A. C. Brown}
\author{H. W. van der Hart}
\affiliation{Centre for Theoretical Atomic, Molecular and Optical Physics,
Queen's University Belfast, Belfast BT7 1NN,
UK}
\date{\today}

\begin{abstract}

We present {\it ab initio} studies of photoelectron spectra for above
threshold detachment (ATD) of F$^-$ anions in short, 1300~nm and 1800~nm
laser pulses.  We identify and assess the importance of electron rescattering in
strong-field photodetachment of a negative ion through comparison with an
analytic, Keldysh-type approach, demonstrating the capability of {\it ab-initio}
computation in the challenging near-IR regime.  We further assess the influence of the strong electron
correlation on the photodetachment.

\end{abstract}

\maketitle



Electron rescattering is one of the fundamental processes occuring in the
interaction between matter and intense light fields \cite{attosecond_review}.
The mechanism is a critical part of 
the well known three-step or recollision
model for high harmonic generation (HHG) or strong field double ionisation. According
to the
model an electron is first ionised, then driven by a strong laser
field, before recolliding with the parent ion, either recombining, leading to
HHG \cite{Co,lewenstein}, or rescattering, leading to high-energy electron emission \cite{hugo_DI,ivanov_DI}, or non-sequential double
ionisation \cite{NSDI}.

Electron rescattering also
encodes structural information about the residual ion into the 
wavepacket of the ejected electron and can thus be exploited as an experimental
probe of the
structure of the parent ion \cite{attosecond_review}. The technique is
especially sensitive as the current density of a recolliding electron wavepacket
exceeds that of conventional electron sources by several orders of magnitude
\cite{recollision_electron_current}. Furthermore, the inherently subcycle and
phase-locked nature of the recollision process gives access to electron dynamics
on the attosecond scale, via information embedded in the photoelectron spectrum
\cite{recollision_molecular_movies,attosecond_2014}.

One of the open questions in strong-field science concerns the importance of electron
rescattering for negative ions.
Significant progress has been made in understanding and controlling the equivalent process
in neutral atoms and positive ions \cite{stereo_ATI_review}, but above-threshold detachment (ATD) presents a
different challenge. The small binding energy allows detachment 
at low intensities. Hence to reach significant recollision energies,
near-infrared (NIR) laser
fields are required. In addition, the absence of the Coulomb potential makes it
easier for the electron wavepacket to spread out, reducing the effect of
rescattering \cite{hugo_DI,ivanov_DI}. While evidence for rescattering from negative ions
has been found experimentally \cite{J}, no verification has yet been provided from {\it
ab initio} theory.  A theoretical approach, based on first order correction to
the strong field approximation, was able to reproduce experimental results from
Br$^-$ and F$^-$, \cite{beck}
but a more recent study, using a numerical solution of the
time-dependent Schr\"odinger equation (TDSE), found
``no qualitative evidence of rescattering" for H$^-$ \cite{Ph}.  In this report
we demonstrate that {\it ab-initio} theory can be used to investigate
rescattering in the NIR regime.

\begin{figure*}[t]
\centering
\includegraphics[width=\linewidth ]{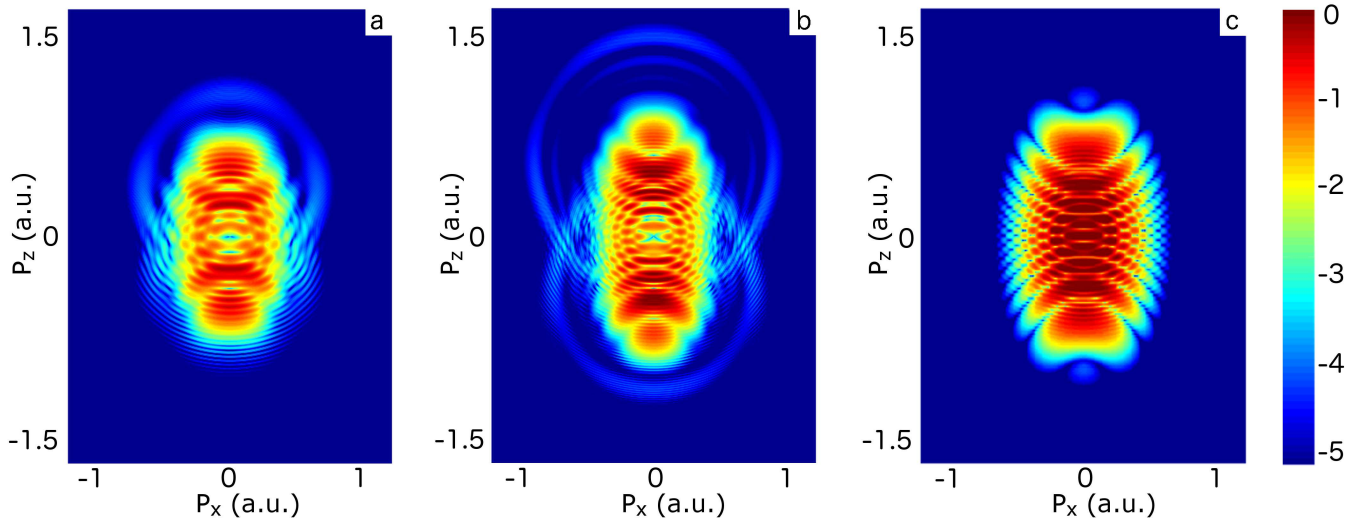}
\caption{(Color online) Electron momentum distributions for ATD of F$^{-}$ in
the $k_x-k_z$ plane for a
(a) 1300~nm (b/c) 1800~nm laser pulse of intensity $1.3 \times 10^{13}$ W/cm$^{2}$.
(a) and (b) are calculated using the RMT approach, while (c) is obtained using
the KTA method \cite{F}. \label{fig:2Dspectra}}
\end{figure*}

An additional complication in the description of negative ions is the much
larger influence of dielectronic-repulsion. Several approximate methods
have been employed to model photodetachment from complex negative ions \cite{F,Ph}, but
these methods are limited in their description of electron repulsion.
A previous study found that 
multiphoton detachment rates for F$^-$ are affected
substantially by the inclusion of correlation effects \cite{Hugo1}. The accurate
description of dielectronic repulsion may prove especially important in
rescattering calculations for negative ions, as the neutral core means the process will be
mediated entirely by short-range effects rather than the long-range potential of
a charged core.

In this letter we address these fundamental questions by applying R-matrix
theory with time-dependence (RMT) to study strong-field dynamics of F$^-$ in
NIR laser pulses. 
The RMT method is an \textit{ab initio}
method for solving the TDSE for
multielectron atomic systems in intense, short laser pulses. As with all R-matrix methods, it
employs a division of space, whereby electron exchange
effects are fully described in an inner-region close to the nucleus, while far
from the nucleus, a single, ejected electron moves in the long-range multipole
potential of the core.
Although several other time-dependent R-matrix methods have emerged in
recent years \cite{van1,Gu,Ly1}, RMT exhibits
orders-of-magnitude improvements in efficiency, primarily because it employs
finite-difference (FD) techniques to model the one-electron wave function in the
outer region. RMT merges the outer region FD model with a 
B-Spline-based, R-matrix basis set for the multielectron inner region, allowing
efficient calculations accounting for atomic structure and correlation effects \cite{Moo11}. 

The R-matrix basis for F$^-$ consists of the neutral F ground
state coupled with an additional electron. We employ two different models for the F atom. These two models
are compared to assess the influence of electron correlation.
The basic model includes only one configuration,
$1s^{2}2s^{2}2p^{5}$, with the $1s$, $2s$ and $2p$ orbitals given by
the Hartree-Fock orbitals for the F ground state \cite{Hugo1,Hugo2}.
The second model includes additional $\overline{3s}, \overline{3p},$ and $\overline{3d}$
pseudoorbitals, \cite{Ni}. This allows us to generate
an accurate wave function expansion for the $1s^22s^22p^5$ $^2P^o$ state from a
configuration interaction calculation including the
$1s^{2}2s^{2}2p^{5}$, $1s^22s2p^53s$, $1s^{2}2s^{2}2p^{4}3p$, $1s^{2}2s^{2}2p^{3}3p^{2}$ and
$1s^{2}2s^{2}2p^{3}3d^{2}$ configurations. 
Model 2 gives a binding energy of 3.421~eV for the $^1S^e$ F ground-state, which
is close to the experimental value of 3.399~eV \cite{F-experiment},
whereas in model 1 we shift the ground-state artificially to 3.421~eV.

We employ 1300~nm and 1800~nm
wavelength fields at a peak intensity of $1.3 \times
10^{13}$ Wcm$^{-2}$. 
The profile comprises two cycles  $\sin^2$ ramp-on
followed by two cycles  $\sin^2$ ramp-off.
In such high intensity, long-wavelength fields, the ejected electron wavepacket can
travel far from the nucleus, and hence, an accurate description of the
wave function is required over an extended region of space. To facilitate this we include angular momenta up to
$L=240$ and propagate the wave function out to a radius of 4265~a$_0$. The time step in the wave function propagation is
0.24 as. 

The laser parameters are chosen to facilitate a comparison with 
results obtained using a Keldysh-Type Approach (KTA) calculation
\cite{GK,short_GK,F}. In this model, the effect of the atomic potential on a
detached electron is neglected. The description of F$^-$ is based on Hartree-Fock
orbitals for F$^-$ in which the long range tail is fitted to the correct binding
energy. The laser field is infinitely long, but with a periodic
envelope which describes a series of short pulses. This allows the analytical
solution of the so-called saddle-point equation, yielding the
electron trajectories in the field. As a consequence of the long laser `pulse',
the field is composed of three distinct photon energies, which is at variance
with our time-dependent R-matrix calculations wherein a spread of photon energies follows
from the isolated short pulse used.



Figure \ref{fig:2Dspectra} shows the 2D electron momentum spectra for ATD from F$^-$
for 1300~nm and
1800~nm fields alongside the KTA spectrum for 1800~nm \cite{F}. The 
$z$-axis is directed along the polarization axis of the laser. The figure
shows results from the basic atomic structure, model 1 of F$^-$ described above,
which allows a closer comparison to the KTA results. 
The F atom has three different possible final states, corresponding to 
$m=0,\pm 1$ where $m$ is the magnetic quantum number.  
The total momentum distribution is
obtained by incoherently summing the contributions from the $m =
0$ and the $m=\pm1$ states.

The ATD momentum distribution in Fig. \ref{fig:2Dspectra} is extensively
detailed. Both the RMT and KTA spectra show the typical structure of rings centered at
zero momentum, with each ring corresponding to the absorption of $N$ photons.
Interference patterns arise because multiple electron trajectories
contribute to each final momentum state \cite{Ph}. Lines connecting the
interference minima are curves in momentum space which satisfy a destructive
interference condition. Curves intersecting the $P_z$-axis correspond to
interference between the well known long- and short-trajectories. Curves
intersecting the $P_x$-axis lead to minima/maxima in the even/odd ATD rings for
perpendicular emission. The two different approaches therefore
demonstrate the same basic physics of photodetachment. However, the features appear sharper in the KTA spectra due to
the well defined photon energy, $\omega$, while for RMT the results are
blurred by uncertainty in $\omega$. Furthermore, the RMT spectrum appears
narrower in the $k_x$ direction.

Figure \ref{fig:2Dspectra} 
shows additional ATD rings for electron momenta greater than the maximum drift
momentum,
$P_z=\mathcal{E}_0/\omega$ (where $\mathcal{E}_0$ is the maximum strength of
the electric field) for both
1300~nm and 1800~nm with the RMT calculations. These extra rings are not present in the KTA model
results.
In the positive $z$ direction, the rings are centred on
$P_z\approx0.55$~a.u.,  and $P_z\approx0.75$~a.u. for 1300~nm and
1800~nm respectively: these values corresponding to the maximum drift momentum
for each wavelength.

\begin{figure}[t]
\centering
\includegraphics[width=\linewidth]{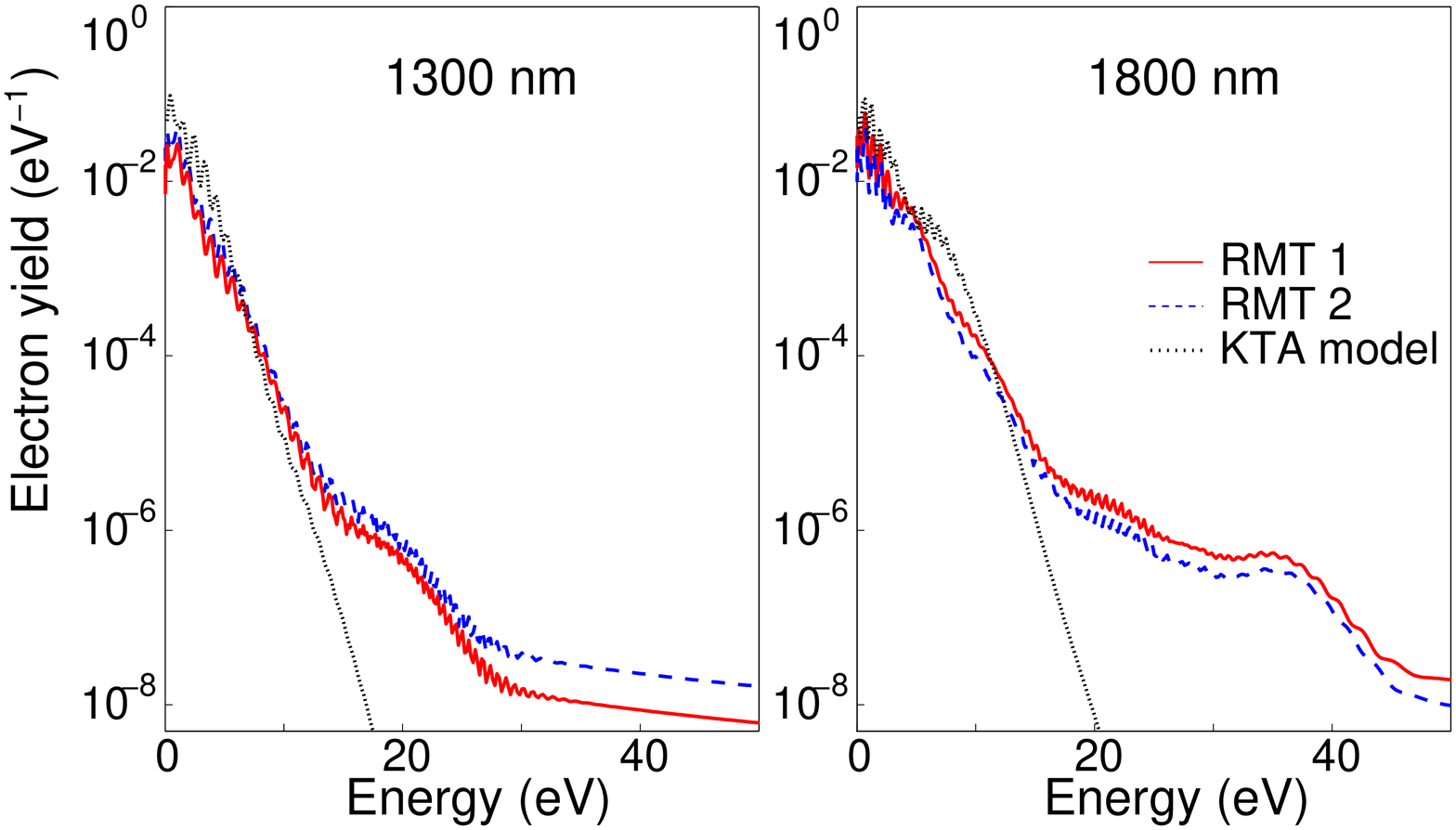}
\caption{(Color online) Photoelectron energy spectra for ATD from F$^-$ by a
  $1.3\times 10^{13}$W/cm$^2$ pulse at 1300~nm and 1800~nm from the present
  approach, model 1 (solid, red line), model 2 (dashed, blue line) and the KTA model
  approach (dotted, black line). The presence of `knee'-like structures in the RMT results is
  a characteristic signature of rescattering. \label{fig:1300PE_log}}
\end{figure}

These additional rings are a clear signature of electron-rescattering.
In the strong field electrons are detached near the maximum of the field each
half cycle as per the three-step model. The electrons are thus `born' into the
field with zero energy, leading to ATD rings centered on zero momentum as in
Fig. \ref{fig:2Dspectra}. Recollision then occurs near the zero-field
intersections, allowing the electron to gain extra drift momentum through its
interaction with the field potential. Rescattered electrons are effectively
`reborn' into the field at the field-zero, gaining a maximum drift momentum of
$P_z=\mathcal{E}_0/\omega$. This can be confirmed with
classical trajectory calculations \cite{varju}. 
The appearance of these rings is proof of the recollision mechanism in negative
ions. Angle resolved spectra showing these rings centered on the drift momentum
go beyond the integrated yields shown in Fig. \ref{fig:1300PE_log} and
correspondingly in \cite{beck}, which show only evidence of high energy
electrons which are ascribed to the recollision mechanism.

The results display a clear asymmetry in the $P_z$ direction, with the
recollision rings in the positive $z$ direction extending to larger momenta than in the negative
$z$ direction. This is due to the short pulse profile we employ. The highest
energy electrons are accelerated by the single, peak-intensity cycle of the laser
pulse and hence are only emitted in the positive direction.  Moreover, in the
negative $z$ direction, there is a pronounced interference structure in the
recollision rings, arising from two interfering electron trajectories in the ramp-on and
ramp-off of the field. Finally, the centre of these recollision rings is shifted
towards the origin for the negative $z$ direction, because the electron
trajectories in the ramp-on and ramp-off of the field are driven by lower
intensity laser cycles. Thus the centre of the recollision rings in the negative
$z$ direction is given by $0.85\times\mathcal{E}_0/\omega$. We note that the
non-recollision part of the momentum distribution is mediated by the vector potential of
the laser field, which is antisymmetric about zero potential.
Thus this part of the distribution is symmetric with respect to the origin, while
the recollision, mediated by the asymmetric electric field, leads to an
asymmetric spectrum. We would expect the opposite to be true, were the carrier
envelope phase of the laser pulse adjusted accordingly.

A second outstanding question surrounding strong-field dynamics of negative ions
concerns the influence of
electron correlation. Since the outer electron in F$^-$ is loosely bound,
and the main rescattering process occurs very close to the residual atom,
electron correlation may affect both the initial ionization and the
rescattering.
A recent study suggests that while for neutral atoms there are notable
differences between KTA model results and those from the numerical solution of
the TDSE, these differences are negligible for negative ions \cite{Ph}.
The ejected-electron momenta spectra obtained for model 1 and model 2 show
little qualitative difference, but notable, quantitative differences can be
observed when electron correlation is included.

Table \ref{tab:pop} gives the population in the outer region--- a measure of
detachment probability in our calculations--- compared with the detachment
probability from the KTA model for 1300~nm and 1800~nm. The yield in our model 1 calculations is
reduced by a factor of 2 from the KTA model result for 1300~nm, whereas the
difference is only about 20\% for 1800~nm. When using the more
sophisticated model 2 for F$^-$, we find a further reduction in the detachment
probability by one third.
The reason for these differences is not
obvious, although the effective potentials differ. The
binding energy in RMT model 1 is shifted artificially to the experimental
value, while model 2
gives the correct binding energy directly. This implies that the short-range
potential, and therefore the wave function, is described more accurately in model 2. This increase in accuracy
may lead to detachment
yields about a third smaller than model 1, and a
factor of 2--3 smaller than the KTA. Differences in short-range potential
can thus lead to significant differences in detachment yield.

\begin{table}[t]
\caption {The level of ionisation of F$^-$ in 1300~nm and 1800~nm fields
  calculated for different peak intensities. This ionised population is presented 
  for the two models used in RMT, compared to the total detachment probability 
  from \cite{F}}
\begin{tabular*}{\columnwidth}{@{\extracolsep{\fill}}cccc}
\hline
\hline
Intensity & \multicolumn{3}{c}{Detachment probability }\\
W/cm$^2$  & RMT(Model 1)  & RMT(Model 2)    & Ref. \cite{F}        \\
\hline
\multicolumn{4}{c}{1300~nm}\\
\hline
$7.7\times10^{12}$ &  0.018&0.011&0.036\\ 
$1.1\times10^{13}$ &  0.045&0.031&0.090\\ 
$1.3\times10^{13}$ &  0.065&0.044&0.139\\
\hline
\multicolumn{4}{c}{1800~nm}\\
\hline
$7.7\times10^{12}$ & 0.020&0.013&0.023\\  
$1.1\times10^{13}$ & 0.055&0.034&0.066\\  
$1.3\times10^{13}$ & 0.080&0.052&0.106\\ 
\hline
\end{tabular*}
\label{tab:pop}
\end{table}

\begin{figure}[t]
\centering
\includegraphics[width=\linewidth]{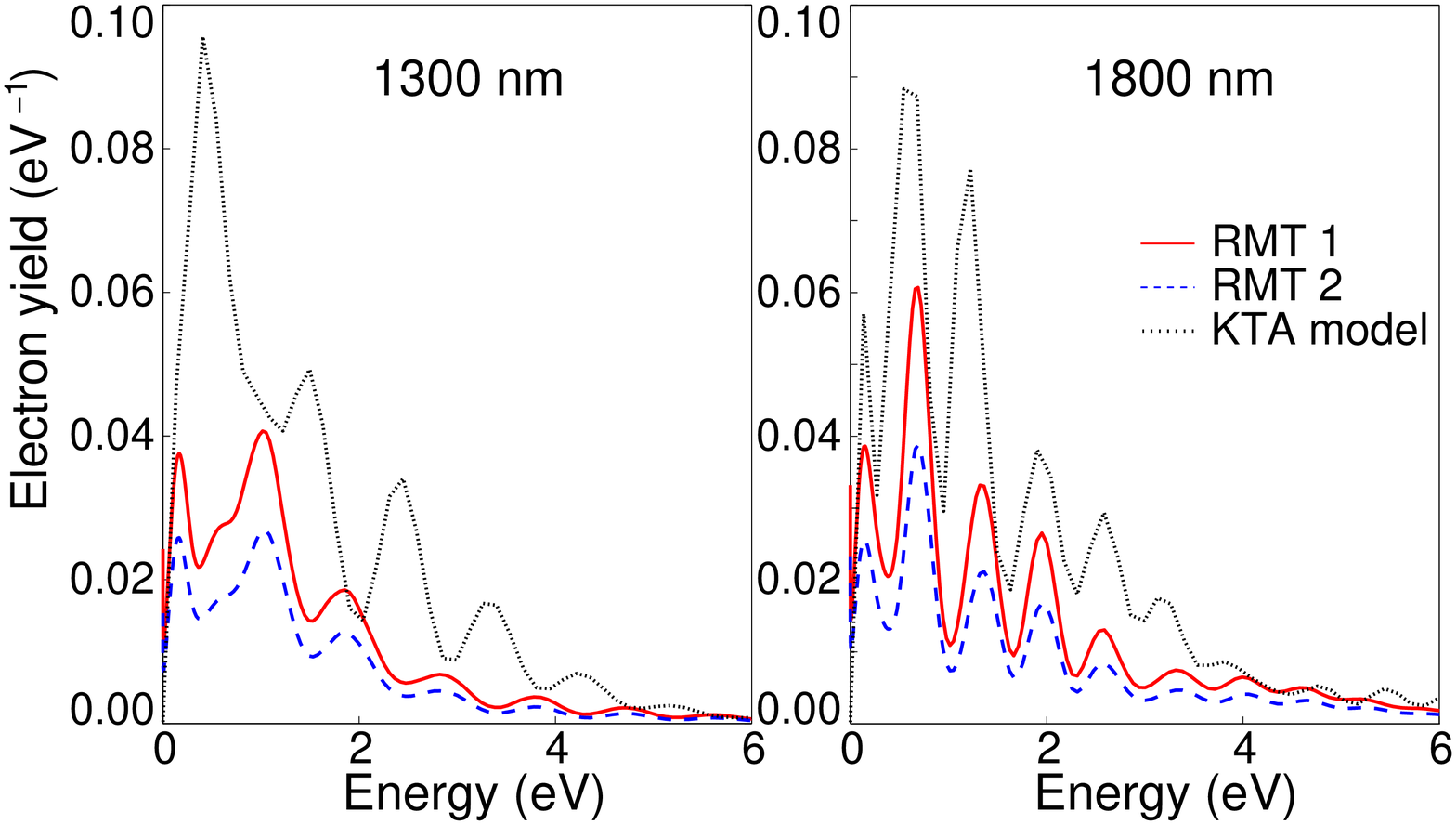}
\caption{(Color online) As Fig. \ref{fig:1300PE_log} but shown for the low energy
region on a linear scale. \label{fig:1300PE_lin}}
\end{figure}

The differences between model 1 and model 2 are most easily identified in the
photelectron energy spectra, which are presented in Figs.
\ref{fig:1300PE_log} and \ref{fig:1300PE_lin} 
alongside results from the KTA model. In all calculations, model 1 shows
a larger electron yield across the full energy range, reflecting the lack of
change in the overall structure of the spectrum. The figures show
that the main effect of electron interaction is an overall reduction
in detachment probability.


Figure \ref{fig:1300PE_log} demonstrates the capability of the RMT approach to
describe electron rescattering. For energies up to about 10~eV, the RMT and KTA
calculations show a very similar behaviour for electron emission, consisting of
a small initial plateau followed by rapid exponential decay. Differences
emerge for energies greater than about 10~eV, with the RMT calculations showing
a plateau at a magnitude of $\sim10^{-5}$--$10^{-6}$~eV$^{-1}$ extending to an energy of about
20~eV for 1300~nm and 35~eV for 1800~nm, corresponding to around 10$U_p$ in each
case. This plateau is absent in the
KTA calculations. Since the KTA approach does not account for rescattering we
can identify this as the source of the plateau. The yields
obtained in model 1 and model 2 are very similar, with those obtained in model 2
lying about one third lower than for model 1.

Figure \ref{fig:1300PE_lin} shows the low energy region of the photoelectron
spectra, which display
peaks characteristic of the multiphoton mechanism: each peak corresponding to
the absorption of an integer number of photons and with energy,  $E =
N\omega-U_{P} -I_{P}$ for some integer $N$, and the ionization energy, $I_p$.
In all cases, the magnitude of the photoelectron peaks is lower for the present
calculations than for the KTA approach and lower for model 2 than for model 1, consistent with the lower
detachment probability discussed above. The qualitative agreement with the KTA
model is excellent for 1800~nm, while, at 1300~nm, the positions of the ATD
peaks differ. The origin of this difference is difficult to assess
unambiguously. However, two factors may play a role. Firstly, the RMT
calculations account for AC-Stark shifts beyond the ponderomotive energy shift,
whereas the KTA model includes the ponderomotive shift only.
Secondly, the RMT method uses a short laser pulse, while the KTA uses an
infinitely long field with a periodic envelope. This leads to differences in the
photon energy and consequently the ponderomotive shift, making it difficult to
determine the correct position of the photoelectron peaks.
It will therefore be interesting to compare
these results with experimental spectra, although, as far as we are aware, only on-axis emission has been
measured for the present laser parameters \cite{ATD8}.

In conclusion, we have demonstrated the capabilitiy of {\it ab initio} theory to
study ATD of
F$^-$ anions in computationally challenging NIR laser pulses with a full description of multielectron
effects. Through comparison with a KTA model we have identified the importance
of the recollision
mechanism in the electron momentum distributions. High energy rings in the angle
resolved photoelectron spectra are the first theoretical verification of rescattering from
negative ions. The integrated photoelectron
energy spectra shows further clear evidence of rescattering, with a knee-like,
plateau structure extending to energies of $10U_p$. 
Although the yield in these rescattering channels is small--
on the order of .02~\% and .07~\% of the total yield for 1300nm and 1800nm
respectively-- it is clear evidence of rescattering.

The description of multielectron atoms and ions in short, intense NIR pulses
is a significant computational challenge, but holds substantial
promise for elucidating new physics, mediated by multielectron interaction.
The RMT method has already been applied to other strong field phenomena,
including high-harmonic generation \cite{Has14a}, but the accurate and efficient
determination of the multielectron wave function over such a large region of space is a first for
calculations of this type. Indeed, the present calculation of ATD in negative
ions provides an extremely
sensitive test of the wave function accuracy. It will be interesting now to
compare theoretical single-atom results of the RMT approach with the findings of
state-of-the-art experimental techniques in strong-field physics and model
approaches \cite{beck}.

\begin{acknowledgments}
OH acknowledges financial support from the University of Jordan. HWH acknowledges
financial support from the UK EPSRC under grant no. EP/G055416/1 and the EU Initial
Training Network CORINF. SL is funded by DEL-NI under the programme for
government.  This work used the
ARCHER UK National Supercomputing Service (\url{http://www.archer.ac.uk}).
The data used in this paper may be accessed via
\url{pure.qub.ac.uk/portal/en/datasets/search.html}.
\end{acknowledgments}

\bibliography{/users/abrown41/Documents/pub/F-ATD/mybib}

\end{document}